\documentclass[acmsmall]{acmart}

\usepackage{amsmath}
\usepackage{url}
\usepackage{tikz-cd}
\tikzcdset{arrow style=math font}

\newcommand\hole{\vphantom{M}}

\DeclareMathOperator{\num}{\mathtt{num}}
\DeclareMathOperator{\str}{\mathtt{str}}
\DeclareMathOperator{\bool}{\mathtt{bool}}
\DeclareMathOperator{\del}{\mathtt{del}}

\DeclareMathOperator{\Ins}{\mathtt{Ins}}
\DeclareMathOperator{\Conv}{\mathtt{Conv}}
\DeclareMathOperator{\Move}{\mathtt{Move}}
\DeclareMathOperator{\Id}{\mathtt{Id}}
\DeclareMathOperator{\project}{\mathtt{project}}
\DeclareMathOperator{\retract}{\mathtt{retract}}
\DeclareMathOperator{\translate}{\mathtt{translate}}

\theoremstyle{acmplain}
\newtheorem{proposition}{Proposition}

\AtBeginDocument{%
  \providecommand\BibTeX{{%
    \normalfont B\kern-0.5em{\scshape i\kern-0.25em b}\kern-0.8em\TeX}}}

\setcopyright{rightsretained}
\copyrightyear{2021}
\acmYear{2021}
\acmDOI{}
\acmISBN{}

\acmConference[HATRA'21]{Human Aspects of Types and Reasoning Assistants}{October 19, 2021}{Chicago, USA}

\begin{document}

\title{Typed Image-based Programming with Structure Editing}

\author{Jonathan Edwards}
\email{jonathanmedwards@gmail.com}
\orcid{0000-0003-1958-79672}
\affiliation{%
  \institution{independent}
  \city{San Diego}
  \country{US}
}

\author{Tomas Petricek}
\email{T.Petricek@kent.ac.uk}
\affiliation{%
  \institution{University of Kent}
  \city{Canterbury}
  \country{UK}
}


\begin{abstract}

  Many beloved programming systems are \textit{image-based}: self-contained worlds that persist both code and data in a single file. Examples include Smalltalk, LISP, HyperCard, Flash, and spreadsheets.
  Image-based programming avoids much of the complexity of modern programming technology stacks and encourages more casual and exploratory programming.
  However conventional file-based programming has better support for collaboration and deployment.
  These problems have been blamed for the limited commercial success of Smalltalk.
  We propose to enable collaboration in image-based programming via types and structure editing.

  We focus on the problem of schema change on persistent data.
  We turn to static types, which paradoxically require more schema change but also provide a mechanism to express and execute those changes.
  To determine those changes we turn to structure editing, so that we can capture changes in type definitions with sufficient fidelity to automatically adapt the data to suit.
  We conjecture that typical schema changes can be handled through structure editing of static types.

  That positions us to tackle collaboration with what could be called version control for structure editing. We present a theory realizing this idea, which is our main technical contribution. While we focus here on editing types, if we can extend the approach to cover the entire programming experience then it would offer a new way to collaborate in image-based programming.

\end{abstract}





\maketitle

\section{Introduction}

Our elders tell stories of the golden age of programming in Smalltalk and LISP machines. These self-contained programming worlds use one language throughout, with all code and data persisted in an \textit{image}\cite{Sandewall78, Goldberg80}. Image-based programming can be much simpler than our modern stack of technologies, each with their own specialized languages and data models. In an image there is no need for a distinct operating system, a file system, a shell language, a build language, a database, and none of the adapters between them. One language within one development environment is used for everything. Image-based programming is thought to encourage \textit{exploratory programming}\cite{Trenouth91, Kery17}. Other beloved programming systems are also forms of image-based programming: HyperCard\cite{Goodman87}, Flash, and spreadsheets are all self-contained programming worlds persisting code and data together into a programmable document -- an image by another name.

Unfortunately collaboration and deployment is harder with image-based programming than with conventional file-based programming. These problems have been blamed for the limited commercial success of Smalltalk\cite{Bracha.history}. Collaboration in file-based programming is done with \textit{version control systems} like git\cite{ProGit}. But adapting that approach to images, like Smalltalk Envy\cite{Envy, Hammant17}, is complicated and clashes with the native style of development. After all, it would be unsatisfying to solve the collaboration problem by importing the complexity of file-based programming.

To develop a new approach to collaborating with images we focus on a propitious special case: \textit{schema change}, which is the problem of adapting persistent data to match code changes.
Version control systems cannot even comprehend schema change, for data and semantics is outside their purview. It is left as a problem for other tools, like Rails Migrations\cite{RailsMigrations}, that must be separately coded and then manually coordinated with code changes.
From the perspective of image-based programming, where code and data live together harmoniously, it is only natural to expect that collaboration tools should coordinate changes to both code and data.
Schema change also serves as a useful entry point because it arises even in solo image programming, allowing us to start in that simpler context.


We use static types.
Dynamically typed systems like Smalltalk can handle simple schema changes, such as adding a field to a class, but not much more. General purpose solutions as in Gemstone\cite{Gemstone} can be complex and code-intensive.

We claim that, ironically, static typing simplifies schema change. In a dynamically typed language any field can hold any value, so who is to say that value must change? It's left up to hand-written code. Putting a static type on a field is declarative: it forces a change to the value when the type changes, and tell us the acceptable range of results. The problem is reduced to inferring the appropriate data changes from a change to the type definition. That is a problem more amenable to tooling. To simplify the problem we will start with a clean slate and defer the complexities of retrofitting our approach into existing systems.

\subsection{Schema Change with Abstract Structure Editing}

Consider editing a record type to rename and reorder a field. If, like version control systems, we look only at the textual changes to the definition then we can't distinguish that case from deleting the field and inserting a new one with a new name. But that distinction matters for whether the field's data should be moved or deleted. Text differencing loses too much information.

To determine schema changes we use \textit{structure editing}\cite{Sandewall78, Teitelbaum81, Barstow1984}.
Instead of defining types with textual syntax we store them as structured values living in the image, and directly manipulate them with a GUI that performs high-level edit operations. For the example of a record type there would be distinct edit operations to insert, delete, move, and rename fields.
By capturing type edits at this fine level of granularity we can perform the correct schema changes to the data, knowing to move and rename it rather than delete it.

Structure editing has been justly criticized for lacking the fluidity of text editing. But note that many image-based programming systems already use structure editing to some extent, editing text only at the finer levels of structure like Smalltalk methods and spreadsheet cells.
In this paper we abstract away from the user interface issues of structure editing and focus on the semantics: what are the edits, how do they work, and how are they managed? A GUI structure editor provides affordances for these edits, but they might also be surfaced in a vi-style editor or a REPL or an API. We defer such interfaces to future work.

We conjecture that typical schema changes can be handled through structure editing of static types.
Given that, the problem of collaborative schema change is reduced to the problem of collaborative structure editing.
That immediately suggests collaborative editing systems like Google Docs, which uses the theory of Operational Transformation (OT)\cite{Ellis89, Ressel96}.
However that isn't the kind of collaboration we need: rather than synchronizing distributed replicas of the same image, we need to support images that are long-lived variants of each other, display their differences, and then allow selected differences to be migrated between them.
What we want is a workflow like textual version control, but for structure editing. We call this \textbf{version control for structure editing}.
To provide this we adapt the theory of OT to the requirements of version control. That is the main technical contribution of this paper, which we now present.

\section{Version Control for Structure Editing}

In this section we present the theory of our approach on the simplest possible domain: a single tuple of atomic types. The type of a \textit{document} (our preferred term for an image) is a tuple of atomic types $(T_1,\ldots T_n)$ where $T_i \in \{\num, \str, \bool, \del\}$. The type $\del$ marks a term as deleted but retains it as a \textit{tombstone}\cite{Oster06}. We define a minimal set of edit operations:

\begin{center}
  $
  \begin{array}{ll}
  \Ins[i, t] &\text{Insert type $t$ at index $i$, shifting indexes $\geq i$ right.}\\
  \Conv[i, t] &\text{Convert index $i$ to have type $t$.}\\
  \Move[i, j] & i \neq j. \text{ Set index $i$ to have the type and value of index $j$ and delete $j$.}\\
  \Id &\text{Does nothing.}\\
  \end{array}
  $
  \end{center}

In section \ref{semantics} we present a precise semantics of edits including values. Until then we simplify the discussion by focusing only on changes to types.

\begin{figure}[t]
\begin{minipage}{1in}
\center
\begin{tikzcd}
A \ar[d, "\mathit{pre}"] \ar[r, "\mathit{diff}"]
&B \ar[d, swap, dashed, "\mathit{post}"]\\
A' \ar[r, dashed, swap, "\mathit{adjust}"]
&B'
\end{tikzcd}
1.~project
\end{minipage}
\quad
\begin{minipage}{1in}
\center
\begin{tikzcd}
A \ar[d, dashed, "\mathit{pre}"] \ar[r, "\mathit{diff}"]
&B \ar[d, swap, "\mathit{post}"]\\
A' \ar[r, dashed, swap, "\mathit{adjust}"]
&B'
\end{tikzcd}
2.~retract
\end{minipage}
\quad
\begin{minipage}{1.2in}
\center
\begin{tikzcd}
A \ar[d, "\mathit{pre}"] \ar[r, "\mathit{diff}"]
&B \ar[d, swap, "\mathit{post}"]\\
A' \ar[r, swap, "\mathit{adjust}"]
&B'
\end{tikzcd}
3.~project \& retract
\end{minipage}
\quad
\begin{minipage}{1.3in}
\center
\begin{tikzcd}[column sep=large]
() \ar[d, dashed, "/" marking] \ar[r, "{\Ins[1, \str]}"]
&(\str) \ar[d, swap, "{\Conv[1, \num]}"]\\
\hole \ar[r, dashed, "/" marking]
&(\num)
\end{tikzcd}
4.~no retraction
\end{minipage}
\caption{Commutative diagrams for $\project$ and $\retract$ operations.}
\Description[project and retract diagrams]{Commutative diagrams for project and retract operations.}
\label{fig:diagrams}
\end{figure}
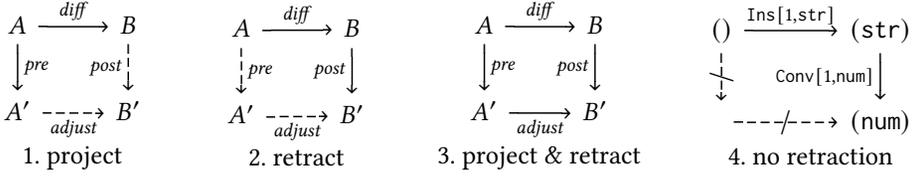

Version control is about tracking the differences between variant documents, and migrating changes between them. The simplest situation is where document $B$ is produced from document $A$ with a single edit \textit{diff}.
A change is then made in the original document $A$ and we want to be able to migrate this change into the derived document $B$. More formally, say that we have an edit \textit{pre} that modifies $A$ into $A'$. We want to determine the edit \textit{post} modifying $B$ (into $B'$) that ``does the same thing''
\footnote{The rules in Appendix \ref{rules} define what we mean by ``doing the same thing'', but we don't know how to define this requirement outside the theory.}.
Since we are tracking the differences between documents we also need to determine the edit \textit{adjust} that is the new difference between the modified variants $A'$ and $B'$. The function $\project$ produces \textit{post} and \textit{adjust} subject to this constraint:

\begin{proposition}
  $
  \project(\mathit{pre}, \mathit{diff}) = (\mathit{post}, \mathit{adjust})
  \implies  \mathit{post} \circ \mathit{diff}
  = \mathit{adjust} \circ \mathit{pre}
  $
  \end{proposition}
which is expressed by the commutative diagram 1 in Figure~\ref{fig:diagrams}. The solid arrows indicate the arguments to $\project$ and the dashed arrows are the results. Note that unlike in Category Theory, our commutative diagrams are oriented, so rotation or reflection changes their meaning.

Inversely we can start with \textit{post} and determine \textit{pre} and \textit{adjust} with the function $\retract$ in diagram 2 in Figure~\ref{fig:diagrams} above that expresses the constraint:

\begin{proposition}
$
\retract(\mathit{post}, \mathit{diff}) = (\mathit{pre}, \mathit{adjust})
\implies  \mathit{post} \circ \mathit{diff}
= \mathit{adjust} \circ \mathit{pre}
$
\end{proposition}

Projection and retraction are often (but not always) inverses. When they are we state that with Diagram 3 above using all solid arrows.

Retraction is not always possible. Diagram 4 above shows that
$\Conv[1, \num]$ cannot be retracted through $\Ins[1, \str]$, which created the term being converted. There is no edit that would do the same thing as the $\Conv$. We say that the $\Conv$ edit \textit{depends upon} the $\Ins$ edit. Diagram 4 uses crossed-out arrows to indicate an impossible retraction.

We will walk through examples to explain the essential properties of projection and retraction. Their full definition is in Appendix \ref{rules}. To start with, the $\Ins$ and $\Move$ operations transform the location of other edits as would be expected:

\begin{center}
$
\begin{tikzcd}[column sep=huge]
(\num) \ar[r, "{\Ins[1, \bool]}"] \ar[d, "{\Conv[1, \str]}"]
&(\bool, \num) \ar[d, "{\Conv[2, \str]}"]\\
(\str) \ar[r, swap, "{\Ins[1, \bool]}"]
&(\bool, \str)\\
\end{tikzcd}
$ \hspace{30pt}$
\begin{tikzcd}[column sep=huge]
(\str, \num) \ar[r, "{\Move[1, 2]}"] \ar[d, "{\Conv[2, \bool]}"]
&(\num, \del) \ar[d, "{\Conv[1, \bool]}"]\\
(\str, \bool) \ar[r, swap, "{\Move[1, 2]}"]
&(\bool, \del)\\
\end{tikzcd}
$
\end{center}

\vskip-1em
An important case is overriding of conflicting edits:

\begin{center}
\begin{tikzcd}[column sep = large]
(\bool) \ar[d, "{\Conv[1, \num]}"] \ar[r, "{\Conv[1, \str]}"]
&(\str) \ar[d, "{\Conv[1, \num]}"] \\
(\num) \ar[r, swap, "\Id"]
&(\num)\\
\end{tikzcd}
\end{center}

\vskip-1em
As in the above example when $\Conv$ operations at the same location conflict then the top edit is overridden so that the left and right edits ``do the same thing''. The top edit gets ``grounded out'' by setting the bottom edit to $\Id$.\footnote{This remains valid when we also consider values because, as discussed in section \ref{semantics}, the $\Conv$ operation is lossless.}

This example shows that $\project$ and $\retract$ are asymmetric, meaning we sometimes cannot flip their diagrams along the diagonal (as in a matrix transpose). This asymmetry expresses that our system does not automatically resolve conflicts, but lets the user chose among possible resolutions (see \ref{migration}). OT obtains symmetry by having timestamps preordain how to resolve conflicts.

Another important case is when duplicate operations ``cancel out'': if the left or right edits duplicate the top one then they make no difference when migrated to the other side, so ``doing the same thing'' demands they ground out to $\Id$. Note how in this case $\project$ and $\retract$ are not inverses:

$\project$: $
\begin{tikzcd}[column sep = huge]
(\str) \ar[d, "{\Conv[1, \num]}"] \ar[r, "{\Conv[1, \num]}"]
&(\num) \ar[d, dashed, "\Id"]\\
(\num) \ar[r, dashed, swap, "\Id"]
&(\num)
\end{tikzcd}
$ \hspace{50pt} $\retract$: $
\begin{tikzcd}[column sep = huge]
(\str) \ar[d, dashed, "\Id"] \ar[r, "{\Conv[1, \num]}"]
&(\num) \ar[d, "{\Conv[1, \num]}"]\\
(\str) \ar[r, dashed, swap, "{\Conv[1, \num]}"]
&(\num)
\end{tikzcd}
$

However except for such cases where $\project$ and $\retract$ ground out, they are inverses. In other words, if an edit ``makes it through'' in one direction, it will return back again restored.

\begin{proposition}
\begin{center}

$\project(\mathit{pre}, \mathit{diff}) = (\mathit{post}, \mathit{adjust})
\land
\mathit{post} \neq \Id
\implies
\retract(\mathit{post}, \mathit{diff}) = (\mathit{pre}, \mathit{adjust})
$

$\retract(\mathit{post}, \mathit{diff}) = (\mathit{pre}, \mathit{adjust})
\land
\mathit{pre} \neq \Id
\implies
\project(\mathit{pre}, \mathit{diff}) = (\mathit{post}, \mathit{adjust})
$

$\project(\Id, \mathit{diff})
= \retract(\Id, \mathit{diff})
= (\Id, \mathit{diff})
$
\end{center}
\end{proposition}

\subsection{Differencing}

The point of version control is to determine the differences between two variants, and migrate selected differences between them. Standard version control is based on heuristic comparison of the variants in reference to an ancestral common state\cite{Khanna07}. We extend the standard approach in two ways. Firstly, we base our comparisons on the actual history of edits in order to be precise enough to perform schema changes.
Secondly, we do not depend on finding a common ancestor. We don't want to require variants to all live in a single shared repository in which to search for an ancestor. Ancestral comparison also inaccurately identifies as differences those changes that have been duplicated in both descendants (perhaps by \textit{cherry picking}\cite{ProGit}).

The point of the ancestor is to localize changes to one document or the other, for example did one document insert something or did the other one delete it? But since we are recording the edit history of each document we already have all the information we need. We can derive the ``optimal ancestor'' which we call the \textit{agreement} $A\&B$ of documents $A$ and $B$. This derivation also produces two sequences of edits called \textit{differences} that reconstruct $A$ and $B$ from $A\&B$. The agreement is optimal in the sense that the differences are minimized, eliminating duplication.

As a starting example, consider the situation where $A$ and $B$ started as $(\num)$ but $A$ inserted a $\bool$ term and $B$ changed the $\num$ term into a $\str$. We diagram the agreement and differences like this:

\begin{center}
\begin{tikzcd}[column sep = huge]
A=(\bool, \num)
&A\&B = (\num) \ar[l, swap, "{\Ins[1, \bool]}"] \ar[r, "{\Conv[1, \str]}"]
&B=(\str)\\
\end{tikzcd}
\end{center}

\vskip-1em
Now say that the user edits $A$ to change the second term $\num$ to be $\str$, as in $B$. In order to adjust the differences we $\translate$ the edit
to $A$ by retracting it backwards to $A\&B$ and then projecting it forwards to $B$, resulting in the diagram:

\begin{center}
\begin{tikzcd}[column sep = huge]
A=(\bool, \num) \ar[d, "{\Conv(2, \str)}"]
&A\&B = (\num) \ar[l, swap, "{\Ins[1, \bool]}"] \ar[d, dashed, "{\Conv(1, \str)}"] \ar[r, "{\Conv[1, \str]}"]
&B=(\str) \ar[d, dashed, "\Id"]\\
A'=(\bool, \str)
&A'\&B' = (\str) \ar[l, dashed, "{\Ins[1, \bool]}"] \ar[r, dashed, swap, "\Id"]
&B'=(\str)\\
\end{tikzcd}
\end{center}

\vskip-1em
Note that in the above diagram the left square is a retraction which has been mirrored to go from left to right. The right-most downward arrow, from $B$ to $B'$, is the translation that ``does the same thing'' as the edit to $A$. It is $\Id$, which means that the original edit doesn't make any difference to $B$ (because $B$ had already made the same change). Therefore the edit is ``absorbed'' into the agreement, as shown in the bottom row of the diagram. Note that $B$ has not changed but the agreement has, and there is no longer any difference between them.

But often edits do make a difference when translated, for example if $A$ had changed the $\num$ to $\bool$, which is different from the change $B$ made:

\begin{center}
\begin{tikzcd}[column sep = huge]
A=(\bool, \num) \ar[d, "{\Conv[2, \bool]}"]
&A\&B = (\num) \ar[l, swap, "{\Ins[1, \bool]}"] \ar[d, dashed, "{\Conv[1, \bool]}"] \ar[r, "{\Conv[1, \str]}"]
&B=(\str) \ar[d, dashed, "{\Conv[1, \bool]}"]\\
(\bool, \bool)
&(\bool) \ar[l, dashed, "{\Ins[1, \bool]}"] \ar[r, dashed, swap, "\Id"]
&(\bool)\\
\end{tikzcd}
\end{center}

\vskip-1em
Here the right-most downwards arrow is not $\Id$, and therefore the original edit is in fact increasing the differences between $A$ and $B$, unlike the previous example. Therefore we do not adjust the agreement -- instead we just append the edit to the differences of $A$ to create $A'$, extending the top row of the previous diagram leftward, and leaving $B$ and the agreement unchanged, diagrammed thusly:

\begin{center}
\begin{tikzcd}[column sep = huge]
A'=(\bool, \bool) &(\bool, \num) \ar[l, swap, "{\Conv[2, \bool]}"] &A'\&B = (\num) \ar[l, swap, "{\Ins[1, \bool]}"] \ar[r, "{\Conv[1, \str]}"] &B=(\str)\\
\end{tikzcd}
\end{center}

\vskip-1em
Another case to consider is if $A$ was to change the first term of the tuple, which does not exist in $A\&B$ or $B$. In that case, the edit would not retract back to the agreement, and as in the previous example would be appended to the differences of $A$.

To handle the general case of multiple differences we define the function $\translate$ that chains together retracts and projects. If $A$ differs from $A\&B$ with the edit sequence $a_{1\ldots n}$, and $B$ likewise with $b_{1\ldots m}$, and given an edit $\epsilon$ to $A$ we retract it through the $a$ edits in reverse order and then project through the $b$ edits in forwards order. This produces an edit $\delta$ to $B$ and adjusted difference sequences $a'$ and $b'$. The whole $\translate$ is undefined if one of the retractions is undefined, meaning that $\epsilon$ depends on one of the $a$ edits. Formally:

\begin{center}
$
\translate(\epsilon, a_{1\dots n}, b_{1\ldots m})
= (\delta, a'_{1\dots n}, b'_{1\ldots m})
$
\quad
where:

\vskip.5em
\begin{tikzcd}[column sep=large]
A \ar{d}[]{\epsilon_n}[swap]{\epsilon}
&\hole \ar[d, dashed, "\epsilon_{n-1}"] \ar[l, swap, "a_n"]
&\hole \ar[l, phantom, marking, "\cdots"] \ar[d, dashed, "\epsilon_1"]
&A\&B \ar[l, swap, "a_1"] \ar[dashed]{d}[]{\delta_0}[swap]{\epsilon_0} \ar[r, "b_1"]
&\hole \ar[d, dashed, "\delta_1"]
&\hole \ar[l, phantom, marking, "\cdots"] \ar[d, dashed, "\delta_{m-1}"] \ar[r, "b_m"]
&B \ar[dashed]{d}[]{\delta}[swap]{\delta_m} \\
A'
&\hole \ar[l, dashed, "a'_n"]
&\hole \ar[l, phantom, marking, "\cdots"]
&A'\&B' \ar[l, dashed, "a'_1"] \ar[r, dashed, swap, "b'_1"]
&\hole
&\hole \ar[l, phantom, marking, "\cdots"] \ar[r, dashed, swap, "b'_m"]
&B'\\
\end{tikzcd}

\vskip-1.5em
\begin{minipage}{.5in}
\begin{align*}
\epsilon_n &= \epsilon\\
\delta_0 &= \epsilon_0\\
\delta &= \delta_m
\end{align*}
\end{minipage}
\begin{minipage}{3in}
\begin{align*}
(\epsilon_{i-1}, a'_i) &= \retract(\epsilon_i, a_i)
\ \ \text{for}\  i = 1 \ldots n\\
(\delta_i, b'_i) &= \project(\delta_{i-1}, b_i)
\ \ \text{for}\  i = 1 \ldots m\\
\end{align*}
\end{minipage}
\end{center}


Now we can state the general algorithm for difference maintenance. If an edit $\epsilon$ to $A$ translates into $\delta = \Id$ on $B$ then the edit is \textit{absorbed} into the agreement and the differences are adjusted into the bottom line of the diagram (leaving $B$ unchanged). If $\delta \neq \Id$ or is undefined then the edit is appended to the differences of $A$ as shown in this diagram:

\begin{center}
\begin{tikzcd}
A'
&A \ar[l, swap, "\epsilon"]
&\hole \ar[l, swap, "a_n"]
&\hole \ar[l, phantom, marking, "\cdots"]
&A'\&B \ar[l, swap, "a_1"] \ar[r, "b_1"]
&\hole
&\hole \ar[l, phantom, marking, "\cdots"]  \ar[r, "b_m"]
&B\\
\end{tikzcd}
\end{center}

\vskip-1em
The above algorithm incrementally maintains the differences and agreement between two documents as they are edited. Note that it operates on edits and never needs to inspect document states.
It is straightforward to rederive the differences from scratch given the edit history of any two documents from a common ancestral state (or their total histories starting from an empty document). This is done by setting the agreement to the common ancestral state, the differences of $B$ to its history from the ancestor, and then applying each of the edits from the history of $A$. It is correct because:

\begin{proposition}
  \label{prop:differencing}
Given two edit histories from the same starting state, the differences calculated will be the same regardless of how the two histories are interleaved.
\end{proposition}

\subsection{Migration}
\label{migration}

In the previous section we defined the function $\translate$ to derive the differences between two documents. The $\translate$ function is also used to migrate individual differences from one document to the other. Returning to where we left off with the last example:

\begin{center}
  \begin{tikzcd}[column sep=huge]
    A = (\bool, \bool)
    & (\bool, \num) \ar[l, swap, "{\Conv(2, \bool)}"]
    & A\&B = (\num) \ar[l, swap, "{\Ins[1, \bool]}"]  \ar[r, "{\Conv[1, \str]}"]
    & B=(\str)\\
  \end{tikzcd}
\end{center}

\vskip-1em
Say the user wants to migrate the last difference of $A$, which set the second term to $\bool$. We remove that edit from $A$'s differences and then translate it through to $B$, producing this diagram:

\begin{center}
  \begin{tikzcd}[column sep = 1.4cm]
    A = (\bool, \bool)
    & (\bool, \num) \ar[l, swap, "{\Conv(2, \bool)}"] \ar[d, "{\Conv[2, \bool]}"]
    & A\&B = (\num) \ar[l, swap, "{\Ins[1, \bool]}"]
    \ar[d, dashed, "{\Conv[1, \bool]}"] \ar[r, "{\Conv[1, \str]}"]
    & B = (\str) \ar[d, dashed, swap, "{\Conv[1, \bool]}"]\\
    A = \phantom{(\bool, \bool)}
    & (\bool, \bool)
    & A\&B' = (\bool) \ar[l, dashed, "{\Ins[1, \bool]}"]
    \ar[r, dashed, swap, "\Id"]
    & B' = (\bool)\\
  \end{tikzcd}
\end{center}

\vskip-1em
$B$ has now been changed equivalently to the edit we picked to migrate, and the differences have been adjusted to remove the now agreed-upon edit.
Now in this example we conveniently chose to migrate the last difference to $A$. In the general case we extract the subsequent differences and then append them back to the results of the translation, as in this diagram showing how to migrate the difference $a_i$:

\begin{center}
\begin{tikzcd}[column sep=1cm]
  A
  &\hole \ar[l, swap, "a_n"]
  &\hole \ar[l, phantom, marking, "\cdots"]
  &\hole \ar[l, swap, "a_{i+1}"]
  &\hole \ar[d, swap, "a_i"] \ar[l, swap, "a_i"]
  &\hole \ar[l, phantom, marking, "\cdots"] \ar[d, dashed]
  &A\&B \ar[l, swap, "a_1"] \ar[dashed]{d} \ar[r, "b_1"]
  &\hole \ar[d, dashed]
  &\hole \ar[l, phantom, marking, "\cdots"] \ar[d, dashed] \ar[r, "b_m"]
  &B \ar[dashed]{d} \\
  A
  &\hole \ar[l, "a_n"]
  &\hole \ar[l, phantom, marking, "\cdots"]
  &\hole \ar[l, "a_{i+1}"]
  &\hole \ar[l, "\Id"]
  &\hole \ar[l, phantom, marking, "\cdots"]
  &A\&B' \ar[l, dashed, "a'_1"] \ar[r, dashed, swap, "b'_1"]
  &\hole
  &\hole \ar[l, phantom, marking, "\cdots"] \ar[r, dashed, swap, "b'_m"]
  &B'\\
  \end{tikzcd}
\end{center}

Consider what happens if $a_1 = \Conv(1, num)$ and $b_1 = \Conv(1, str)$. Migrating $a_1$ will override $b_1$, setting $b'_1 = \Id$. We call this a \textit{conflict}: a merge where one of the $b$ edits is grounded out to $\Id$. A conflicting migration overrides differences in the opposite doc. Conflict is symmetric: if $a$ conflicts with $b$ then $b$ also conflicts with $a$, so the user has the choice of migrating either one and overriding the other. Note that our concept of a conflict is more precise than the \textit{merge conflict} of textual version control, which means only that changes appear to be occurring on the same text line, and often must be resolved with manual text editing. Many of our edits never conflict (like $\Ins$ and $\Move$), and conflicts that do occur are definite, and are resolved just by choosing which one wins.

\begin{proposition}[Conflict symmetry]
  If migrating $a_i$ conflicts with $b_j$ then migrating $b_j$ will conflict with $a_i$.
\end{proposition}

The key property of migration is that it reduces differences:
\begin{proposition}[Convergence]
  \label{convergence}
  Repeatedly migrating differences between two documents will terminate with them equal.
\end{proposition}
Note however that the resultant document value can vary depending on the chosen order of migration. That order determines which of a pair of conflicting edits will dominate.

It is not possible to migrate a difference when it depends on an earlier difference -- the dependency must be migrated first. In our implementation we provide the option to automatically migrate all transitive dependencies. We also support migrating all differences (equivalent to merging a branch in traditional version control) by migrating each in turn. Because we ``cherry pick'' at the granularity of individual edits the user interface can flexibly cluster migrations, say by time or location.

There is a technical problem with migrating $\Ins$ edits. Proposition \ref{prop:differencing} says we can recompute the differences between two documents from their histories, so the adjusted differences produced by the migrate function must be the same as immediately recalculating them afterwards. This fails when we migrate an $\Ins$ edit, because we consider all inserts to be distinct.
To fix this problem we add a unique identifier to the $\Ins$ operation and we modify the $\project$ function to cancel out inserts with the same identifier. We left this detail out of the prior discussion for the sake of simplicity.

\subsection{Value Semantics}
\label{semantics}

The preceding discussion was simplified by considering only the type definitions of tuples. We will now incorporate values, and provide a precise semantics of the edit operations.

A document is a typed tuple $(v_1\colon T_1 \ldots v_n\colon T_n)$ where $T_i \in \{\num, \str, \bool, \del\}$. The type $\del$ has a single value $\mathtt{null}$.

The edit operations are:

\begin{center}
\begin{scriptsize}
$
\begin{array}{ll}
\Id &\text{Does nothing.}\\
\Ins[i, t, p] &\text{Insert type $t$ at index $i$, shifting indexes $\geq i$ right. $p$ is a unique identifier.}\\
\Conv[i, t] &\text{Convert index $i$ to have type $t$.}\\
\Move[i, j] & i \neq j. \text{ Set index $i$ to have the type and value of index $j$ and delete $j$.}\\
\end{array}
$
\end{scriptsize}
\end{center}

defined as:
\begin{scriptsize}
\begin{align*}
\Id\,(v_1\colon T_1 \ldots v_n\colon T_n)
& = (v_1\colon T_1 \ldots v_n\colon T_n) \\
\Ins[i, t, p]\,(v_1\colon T_1 \ldots v_n\colon T_n)
& = (v_1\colon T_1  \ldots v_{i-1}\colon T_{i-1}, \mathtt{null}\colon t, v_i\colon T_i \ldots v_n\colon T_n) \\
\Conv[i, t]\,(v_1\colon T_1 \ldots v_n\colon T_n)
&= (v_1\colon T_1 \ldots v_{i-1}\colon T_{i-1}, v_i\colon t,
v_{i+1} \colon T_{i+1} \ldots v_n\colon T_n) \\
\Move[i, j]\,(v_1\colon T_1 \ldots v_n\colon T_n)
&= (v'_1\colon T'_i \ldots v'_n\colon T'_n)
\text{ where }
v'_k\colon T'_k =
\begin{cases}
v_j\colon T_j & k = i\\
\mathtt{null}\colon \del & k = j\\
v_k\colon T_k & \text{otherwise}\\
\end{cases}
\end{align*}
\end{scriptsize}

The rules defining projection and retraction are presented in Appendix \ref{rules}.

We allow a document to contain type errors. In the spirit of image-based programming, we handle errors gracefully and allow them to be corrected when convenient. We define a function $\mathtt{conform}$ that takes a document and produces a type-conforming version of it, in which all values either conform to their type or are the special value $\mathtt{error}$. Programs can only see conforming values. This technique is related to the \textit{non-empty holes} of Omar et al.\cite{Hazelnut17, HazelnutLive19}.

The $\Ins$ edit gives a new term the value $\mathtt{null}$, representing an uninitialized value. The $\mathtt{conform}$ function can treat an uninitialized value as an error, or as we do in our implementation, assign a default value based on the type.

The $\mathtt{conform}$ function also handles type conversion. The $\Conv$ edit
just sets the desired type, leaving the existing value in place, which
$\mathtt{conform}$ converts if possible or yields $\mathtt{error}$. By retaining
the unconverted value we also allow conversions to compose losslessly, which
occurs when we migrate conflicting conversions. Lossless conversions are
needed to ensure convergence (Proposition \ref{convergence}).

\section{Prototype Implementation}

Will this theory work in practice? We are building a prototype implementation to find out. We have extended the minimalist theory to a more practical set of datatypes and edit operations. We have also built a GUI to explore how to visually manifest structural version control. We have extended the theory to include:

\begin{enumerate}
  \item Nested records and homogeneous lists, forming a tree.
  \item Edit operations are located at paths within the tree, expressed as sequences of indexes.
  \item $\Move$ carries a subtree across the tree to replace another subtree. This includes the ability to reorder lists.
  \item Edits to the type of a list are iterated over its elements.
  \item A rename operation on record fields.
  \item Differences can be \textit{normalized}, producing a sequence of edits ordered by location in the tree.
  \item (pending) Sum types, to gain parity with standard FP datatypes.
  \item (pending) A wrap operation to replace a location the tree with a record or sum or list containing it. There is an inverse unwrap operation.
\end{enumerate}


Our prototype GUI focuses on viewing the differences between two documents and performing migrations. A central problem of structure editing is to make it as ''fluid`` as text editing. That is not the problem we are working on here, so we edit with a command line. We use infix syntax for edit operations, with the prefixed location formatted as a dotted path of field names and list indexes.

\begin{figure}[h]
  \centering
  \includegraphics[width=\textwidth]{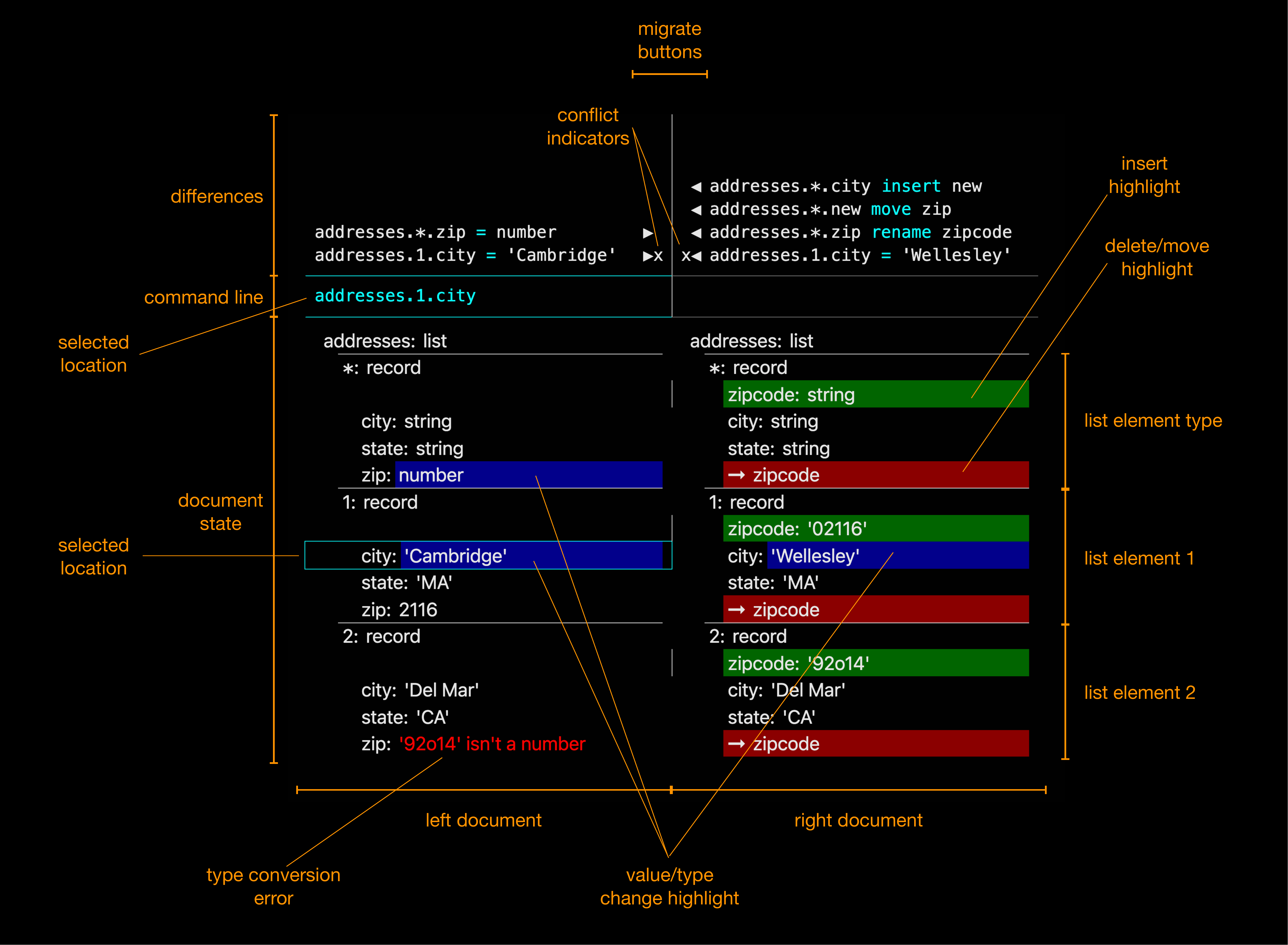}
  \caption{Annotated screenshot of difference view}
  \label{screenshot}
  \Description[screenshot]{Annotated screenshot of difference view}
\end{figure}

Figure \ref{screenshot} shows an annotated screenshot.
A recorded demo can be viewed at
\url{https://vimeo.com/manage/videos/631461226}.

\newpage

\section{Limitations and Opportunities}

This is early stage work that raises more questions than it answers:

\begin{enumerate}
  \item Can direct manipulation editing in a GUI compete with the fluidity of text editing?
  \item Can editing be extended to encompass code, enabling coordinated refactorings across code and data?
  \item Can complex interrelated schema changes be accommodated? Ambler\&Sadalage\cite{ambler2006refactoring} cast many such large scale schema changes as refactorings, which we hope to support as structure edits. Nevertheless in some cases it will still be necessary to write code to convert data. How does such code interact with edit migration?
  \item Can these techniques be retrofitted into existing image-based systems (Smalltalk, LISP), or even text-based ones?
\end{enumerate}

Version control has become a central technology of programming, but we see the chance to enlarge its impact even further:

\begin{enumerate}
  \item Support a larger audience of programmers by providing a simpler and more tractable conceptual model than git \cite{gitless16}.
  \item Images could be utilized like OS containers, so that deployment becomes version control. Deployment was a major problem in the earlier age of image-based programming.
  \item We mentioned at the outset that spreadsheets are an example of image-based programming. We would love to have our approach applied to spreadsheets. Instead of making tweaked copies and copy-pasting between them, we could view their differences and migrate changes as needed.
\end{enumerate}

Our long term vision is to use types and structure editing to bring back the benefits of image-based programming to modern programmers. We hope that we have at least incentivised future research on structure editing by identifying two heretofore overlooked benefits:

\begin{enumerate}
  \item Automatic schema change, and
  \item high-level version control.
\end{enumerate}

\section{Related Work}

SQL DDL\cite{AlterTable} is the standard language for editing relational database schema. It can rename, reorder, and convert the datatype of a column. More complex database conversions can be coded in specialized languages such as Rails Migrations\cite{RailsMigrations}. Schema change has been cast as a form of refactoring\cite{ambler2006refactoring}. Versioning has been built into or added onto databases and other structured data formats\cite{noms, dolt, dbmaestro, daff} and notebooks\cite{Kery17}. The theory of lenses has been applied to schema change on data structures\cite{Foster2007, Cambria}. Closer to our focus on image-based programming is the work on schema change for Object Oriented Databases\cite{Gemstone, Banerjee87}, version control for Smalltalk\cite{Envy}, and Orthogonal Synchronization\cite{Bracha05}.

Research on structure editing has a long history\cite{Sandewall78, Teitelbaum81, Barstow1984}. One focus has been to enable non-textual \textit{projections} of code\cite{intentional, MPS}. Perhaps the most successful structure editor is Scratch\cite{Resnick09}, providing an easy editing experience for beginners free of syntax errors. Scratch and successor \textit{blocks-based} languages rely on drag-and-drop gestures, which motivated our inclusion of the $\Move$ operation.
Recent research seeks to match the fluidity of text editing \cite{Kolling15, Hempel18} and the rigorous semantics of PL theory\cite{Hazelnut17, HazelnutLive19}.

Our work repurposes Operational Transformation (OT) \cite{Ellis89, Ressel96},
which is the theory behind collaborative document editing as in Google Docs. OT
supports collaborative editing of a single document, where the central problem
is the unpredictable arrival order of edits at distributed replicas.
Our need to track and selectively migrate differences between long-lived variants leads to a substantially different and more complex theory.
For one thing, we give users the choice of how and when to resolve conflicts, whereas in OT timestamps preordain resolution.
This introduces asymmetries into our transformations, violating the Symmetry Property of Ressel et al.\cite{Ressel96}.
The notation of commutative diagrams from Category Theory has proven essential to handle the complexity arising, ironically, from the non-commutativity of edits.

The Categorical Theory of Patches \cite{PatchTheory} implemented in Pijul\cite{Pijul} presents a theory of textual version control based on line-oriented text edits called patches. Intriguingly, a merge is modelled as a categorical pushout, diagrammed like our $\project$ function, and our $\retract$ function looks like a pullback. However the asymmetries introduced by overriding conflicts don't satisfy the definitions of pushout and pullback. We don't know what to make of this and welcome any insights from Category Theorists.

Our prior work on Subtext \cite{Subtext05, FirstClassCopyPaste} assigned permanent parent-unique IDs to tree nodes, forming stable paths. However differencing/migrating movement and wrapping/unwrapping operations became complex and confusing. Ultimately we felt forced to adopt an OT-like approach of transforming edit histories in order to have some confidence of correctness. Folk wisdom has it that OT is difficult to get right, but so far we have found the implementation guided by the theory to be robust. Note that unique IDs had to be incorporated into $\Ins$ operations to support Proposition \ref{prop:differencing}, so the status of IDs is still unclear. It would be interesting to see if recent CRDT advances \cite{Kleppmann21} could provide an alternative solution.

\section{Acknowledgements}
We thank Philip Guo, Joel Jakubovic, and the anonymous reviewers for constructive comments.

\bibliographystyle{ACM-Reference-Format}
\bibliography{HATRA21}


\begin{thebibliography}{40}


\ifx \showCODEN    \undefined \def \showCODEN     #1{\unskip}     \fi
\ifx \showDOI      \undefined \def \showDOI       #1{#1}\fi
\ifx \showISBNx    \undefined \def \showISBNx     #1{\unskip}     \fi
\ifx \showISBNxiii \undefined \def \showISBNxiii  #1{\unskip}     \fi
\ifx \showISSN     \undefined \def \showISSN      #1{\unskip}     \fi
\ifx \showLCCN     \undefined \def \showLCCN      #1{\unskip}     \fi
\ifx \shownote     \undefined \def \shownote      #1{#1}          \fi
\ifx \showarticletitle \undefined \def \showarticletitle #1{#1}   \fi
\ifx \showURL      \undefined \def \showURL       {\relax}        \fi
\providecommand\bibfield[2]{#2}
\providecommand\bibinfo[2]{#2}
\providecommand\natexlab[1]{#1}
\providecommand\showeprint[2][]{arXiv:#2}

\bibitem[\protect\citeauthoryear{??}{Alt}{2021}]%
        {AlterTable}
 \bibinfo{year}{2021}\natexlab{}.
\newblock \bibinfo{title}{ALTER TABLE, SQL Language Reference}.
\newblock
\newblock
\urldef\tempurl%
\url{https://docs.oracle.com/en/database/oracle/oracle-database/21/sqlrf/ALTER-TABLE.html#GUID-552E7373-BF93-477D-9DA3-B2C9386F2877}
\showURL{%
\tempurl}


\bibitem[\protect\citeauthoryear{??}{dbm}{2021}]%
        {dbmaestro}
 \bibinfo{year}{2021}\natexlab{}.
\newblock \bibinfo{title}{DBmaestro: DevOps for Database}.
\newblock
\newblock
\urldef\tempurl%
\url{https://db.dbmaestro.com/}
\showURL{%
\tempurl}


\bibitem[\protect\citeauthoryear{??}{dol}{2021}]%
        {dolt}
 \bibinfo{year}{2021}\natexlab{}.
\newblock \bibinfo{title}{Dolt}.
\newblock
\newblock
\urldef\tempurl%
\url{https://www.dolthub.com/}
\showURL{%
\tempurl}


\bibitem[\protect\citeauthoryear{??}{Rai}{2021}]%
        {RailsMigrations}
 \bibinfo{year}{2021}\natexlab{}.
\newblock \bibinfo{title}{Migrations}.
\newblock
\newblock
\urldef\tempurl%
\url{https://guides.rubyonrails.org/v3.2/migrations.html}
\showURL{%
\tempurl}


\bibitem[\protect\citeauthoryear{??}{Pij}{2021}]%
        {Pijul}
 \bibinfo{year}{2021}\natexlab{}.
\newblock \bibinfo{title}{Pijul, a sound and fast distributed version control
  system based on a mathematical theory of asynchronous work}.
\newblock
\newblock
\urldef\tempurl%
\url{https://pijul.org/}
\showURL{%
\tempurl}


\bibitem[\protect\citeauthoryear{Ambler and Sadalage}{Ambler and
  Sadalage}{2006}]%
        {ambler2006refactoring}
\bibfield{author}{\bibinfo{person}{S.W. Ambler} {and} \bibinfo{person}{P.J.
  Sadalage}.} \bibinfo{year}{2006}\natexlab{}.
\newblock \bibinfo{booktitle}{\emph{Refactoring Databases: Evolutionary
  Database Design}}.
\newblock \bibinfo{publisher}{Addison Wesley}, \bibinfo{address}{USA}.
\newblock
\showISBNx{9780321293534}
\showLCCN{2005031959}


\bibitem[\protect\citeauthoryear{Attic-Labs}{Attic-Labs}{2021}]%
        {noms}
\bibfield{author}{\bibinfo{person}{Attic-Labs}.}
  \bibinfo{year}{2021}\natexlab{}.
\newblock \bibinfo{title}{Noms}.
\newblock
\newblock
\urldef\tempurl%
\url{https://github.com/attic-labs/noms}
\showURL{%
\tempurl}


\bibitem[\protect\citeauthoryear{Banerjee, Kim, Kim, and Korth}{Banerjee
  et~al\mbox{.}}{1987}]%
        {Banerjee87}
\bibfield{author}{\bibinfo{person}{Jay Banerjee}, \bibinfo{person}{Won Kim},
  \bibinfo{person}{Hyoung-Joo Kim}, {and} \bibinfo{person}{Henry~F. Korth}.}
  \bibinfo{year}{1987}\natexlab{}.
\newblock \showarticletitle{Semantics and Implementation of Schema Evolution in
  Object-Oriented Databases}.
\newblock \bibinfo{journal}{\emph{SIGMOD Rec.}} \bibinfo{volume}{16},
  \bibinfo{number}{3} (\bibinfo{date}{Dec.} \bibinfo{year}{1987}),
  \bibinfo{pages}{311–322}.
\newblock
\showISSN{0163-5808}
\urldef\tempurl%
\url{https://doi.org/10.1145/38714.38748}
\showDOI{\tempurl}


\bibitem[\protect\citeauthoryear{Barstow, Guty, Shrobe, and Sandewall}{Barstow
  et~al\mbox{.}}{1984}]%
        {Barstow1984}
\bibfield{author}{\bibinfo{person}{D.R. Barstow}, \bibinfo{person}{S.G. Guty},
  \bibinfo{person}{H.E. Shrobe}, {and} \bibinfo{person}{E. Sandewall}.}
  \bibinfo{year}{1984}\natexlab{}.
\newblock \bibinfo{booktitle}{\emph{Interactive Programming Environments}}.
\newblock \bibinfo{publisher}{McGraw-Hill}, \bibinfo{address}{USA}.
\newblock
\showISBNx{9780070038851}
\showLCCN{83013572}


\bibitem[\protect\citeauthoryear{Bracha}{Bracha}{2005}]%
        {Bracha05}
\bibfield{author}{\bibinfo{person}{Gilad Bracha}.}
  \bibinfo{year}{2005}\natexlab{}.
\newblock \bibinfo{title}{Objects as Software Services}.
\newblock
\newblock
\urldef\tempurl%
\url{http://bracha.org/objectsAsSoftwareServices.pdf}
\showURL{%
\tempurl}


\bibitem[\protect\citeauthoryear{Bracha}{Bracha}{2020}]%
        {Bracha.history}
\bibfield{author}{\bibinfo{person}{Gilad Bracha}.}
  \bibinfo{year}{2020}\natexlab{}.
\newblock \bibinfo{title}{Bits of History, Words of Advice}.
\newblock
\newblock
\urldef\tempurl%
\url{https://gbracha.blogspot.com/2020/05/bits-of-history-words-of-advice.html}
\showURL{%
\tempurl}


\bibitem[\protect\citeauthoryear{Chacon and Straub}{Chacon and Straub}{2014}]%
        {ProGit}
\bibfield{author}{\bibinfo{person}{Scott Chacon} {and} \bibinfo{person}{Ben
  Straub}.} \bibinfo{year}{2014}\natexlab{}.
\newblock \bibinfo{booktitle}{\emph{Pro Git} (\bibinfo{edition}{2nd} ed.)}.
\newblock \bibinfo{publisher}{Apress}, \bibinfo{address}{USA}.
\newblock
\showISBNx{1484200772}


\bibitem[\protect\citeauthoryear{Czarnecki and Eisenecker}{Czarnecki and
  Eisenecker}{2000}]%
        {intentional}
\bibfield{author}{\bibinfo{person}{Krzysztof Czarnecki} {and}
  \bibinfo{person}{Ulrich~W. Eisenecker}.} \bibinfo{year}{2000}\natexlab{}.
\newblock \bibinfo{booktitle}{\emph{Generative Programming: Methods, Tools, and
  Applications}}.
\newblock \bibinfo{publisher}{ACM Press/Addison-Wesley Publishing Co.},
  \bibinfo{address}{USA}, Chapter~11.
\newblock
\showISBNx{0201309777}


\bibitem[\protect\citeauthoryear{De~Rosso and Jackson}{De~Rosso and
  Jackson}{2016}]%
        {gitless16}
\bibfield{author}{\bibinfo{person}{Santiago~Perez De~Rosso} {and}
  \bibinfo{person}{Daniel Jackson}.} \bibinfo{year}{2016}\natexlab{}.
\newblock \showarticletitle{Purposes, Concepts, Misfits, and a Redesign of
  Git}.
\newblock \bibinfo{journal}{\emph{SIGPLAN Not.}} \bibinfo{volume}{51},
  \bibinfo{number}{10} (\bibinfo{date}{Oct.} \bibinfo{year}{2016}),
  \bibinfo{pages}{292–310}.
\newblock
\showISSN{0362-1340}
\urldef\tempurl%
\url{https://doi.org/10.1145/3022671.2984018}
\showDOI{\tempurl}


\bibitem[\protect\citeauthoryear{Edwards}{Edwards}{2005}]%
        {Subtext05}
\bibfield{author}{\bibinfo{person}{Jonathan Edwards}.}
  \bibinfo{year}{2005}\natexlab{}.
\newblock \showarticletitle{Subtext: Uncovering the Simplicity of Programming}.
  In \bibinfo{booktitle}{\emph{Proceedings of the 20th Annual ACM SIGPLAN
  Conference on Object-Oriented Programming, Systems, Languages, and
  Applications}} (San Diego, CA, USA) \emph{(\bibinfo{series}{OOPSLA '05})}.
  \bibinfo{publisher}{Association for Computing Machinery},
  \bibinfo{address}{New York, NY, USA}, \bibinfo{pages}{505–518}.
\newblock
\showISBNx{1595930310}
\urldef\tempurl%
\url{http://www.subtext-lang.org/OOPSLA05.pdf}
\showURL{%
\tempurl}


\bibitem[\protect\citeauthoryear{Edwards}{Edwards}{2006}]%
        {FirstClassCopyPaste}
\bibfield{author}{\bibinfo{person}{Jonathan Edwards}.}
  \bibinfo{year}{2006}\natexlab{}.
\newblock \bibinfo{booktitle}{\emph{First Class Copy \& Paste}}.
\newblock \bibinfo{type}{{T}echnical {R}eport} MIT-CSAIL-TR-2006-037.
  \bibinfo{institution}{MIT}.
\newblock
\urldef\tempurl%
\url{http://hdl.handle.net/1721.1/32980}
\showURL{%
\tempurl}


\bibitem[\protect\citeauthoryear{Ellis and Gibbs}{Ellis and Gibbs}{1989}]%
        {Ellis89}
\bibfield{author}{\bibinfo{person}{C.~A. Ellis} {and} \bibinfo{person}{S.~J.
  Gibbs}.} \bibinfo{year}{1989}\natexlab{}.
\newblock \showarticletitle{Concurrency Control in Groupware Systems}. In
  \bibinfo{booktitle}{\emph{Proceedings of the 1989 ACM SIGMOD International
  Conference on Management of Data}} (Portland, Oregon, USA)
  \emph{(\bibinfo{series}{SIGMOD '89})}. \bibinfo{publisher}{Association for
  Computing Machinery}, \bibinfo{address}{New York, NY, USA},
  \bibinfo{pages}{399–407}.
\newblock
\showISBNx{0897913175}
\urldef\tempurl%
\url{https://doi.org/10.1145/67544.66963}
\showDOI{\tempurl}


\bibitem[\protect\citeauthoryear{Fitz}{Fitz}{2021}]%
        {daff}
\bibfield{author}{\bibinfo{person}{Paul Fitz}.}
  \bibinfo{year}{2021}\natexlab{}.
\newblock \bibinfo{title}{daff}.
\newblock
\newblock
\urldef\tempurl%
\url{https://paulfitz.github.io/daff/}
\showURL{%
\tempurl}


\bibitem[\protect\citeauthoryear{Foster, Greenwald, Moore, Pierce, and
  Schmitt}{Foster et~al\mbox{.}}{2007}]%
        {Foster2007}
\bibfield{author}{\bibinfo{person}{J.~Nathan Foster},
  \bibinfo{person}{Michael~B. Greenwald}, \bibinfo{person}{Jonathan~T. Moore},
  \bibinfo{person}{Benjamin~C. Pierce}, {and} \bibinfo{person}{Alan Schmitt}.}
  \bibinfo{year}{2007}\natexlab{}.
\newblock \showarticletitle{Combinators for bidirectional tree transformations:
  {A} linguistic approach to the view-update problem}.
\newblock \bibinfo{journal}{\emph{ACM Transactions on Programming Languages and
  Systems}} \bibinfo{volume}{29}, \bibinfo{number}{3} (\bibinfo{date}{May}
  \bibinfo{year}{2007}), \bibinfo{pages}{17}.
\newblock
\urldef\tempurl%
\url{https://doi.org/citation.cfm?doid=1232420.1232424}
\showDOI{\tempurl}


\bibitem[\protect\citeauthoryear{Goldberg}{Goldberg}{1984}]%
        {Goldberg80}
\bibfield{author}{\bibinfo{person}{Adele Goldberg}.}
  \bibinfo{year}{1984}\natexlab{}.
\newblock \bibinfo{booktitle}{\emph{SMALLTALK-80: The Interactive Programming
  Environment}}.
\newblock \bibinfo{publisher}{Addison-Wesley Longman Publishing Co., Inc.},
  \bibinfo{address}{USA}.
\newblock
\showISBNx{0201113724}


\bibitem[\protect\citeauthoryear{Goodman and Atkinson}{Goodman and
  Atkinson}{1987}]%
        {Goodman87}
\bibfield{author}{\bibinfo{person}{D. Goodman} {and} \bibinfo{person}{W.
  Atkinson}.} \bibinfo{year}{1987}\natexlab{}.
\newblock \bibinfo{booktitle}{\emph{The Complete HyperCard Handbook}}.
\newblock \bibinfo{publisher}{Bantam Books}, \bibinfo{address}{USA}.
\newblock
\showISBNx{9780553343915}
\showLCCN{87201799}


\bibitem[\protect\citeauthoryear{Hammant}{Hammant}{2017}]%
        {Hammant17}
\bibfield{author}{\bibinfo{person}{Paul Hammant}.}
  \bibinfo{year}{2017}\natexlab{}.
\newblock \bibinfo{title}{Smalltalk Envy}.
\newblock
\newblock
\urldef\tempurl%
\url{https://paulhammant.com/2017/09/01/smalltalk-envy/}
\showURL{%
\tempurl}


\bibitem[\protect\citeauthoryear{Hempel, Lubin, Lu, and Chugh}{Hempel
  et~al\mbox{.}}{2018}]%
        {Hempel18}
\bibfield{author}{\bibinfo{person}{Brian Hempel}, \bibinfo{person}{Justin
  Lubin}, \bibinfo{person}{Grace Lu}, {and} \bibinfo{person}{Ravi Chugh}.}
  \bibinfo{year}{2018}\natexlab{}.
\newblock \showarticletitle{Deuce: A Lightweight User Interface for Structured
  Editing}. In \bibinfo{booktitle}{\emph{Proceedings of the 40th International
  Conference on Software Engineering}} (Gothenburg, Sweden)
  \emph{(\bibinfo{series}{ICSE '18})}. \bibinfo{publisher}{Association for
  Computing Machinery}, \bibinfo{address}{New York, NY, USA},
  \bibinfo{pages}{654–664}.
\newblock
\showISBNx{9781450356381}
\urldef\tempurl%
\url{https://doi.org/10.1145/3180155.3180165}
\showDOI{\tempurl}


\bibitem[\protect\citeauthoryear{Kery, Horvath, and Myers}{Kery
  et~al\mbox{.}}{2017}]%
        {Kery17}
\bibfield{author}{\bibinfo{person}{Mary~Beth Kery}, \bibinfo{person}{Amber
  Horvath}, {and} \bibinfo{person}{Brad Myers}.}
  \bibinfo{year}{2017}\natexlab{}.
\newblock \bibinfo{booktitle}{\emph{Variolite: Supporting Exploratory
  Programming by Data Scientists}}.
\newblock \bibinfo{publisher}{Association for Computing Machinery},
  \bibinfo{address}{New York, NY, USA}, \bibinfo{pages}{1265–1276}.
\newblock
\showISBNx{9781450346559}
\urldef\tempurl%
\url{https://doi.org/10.1145/3025453.3025626}
\showURL{%
\tempurl}


\bibitem[\protect\citeauthoryear{Khanna, Kunal, and Pierce}{Khanna
  et~al\mbox{.}}{2007}]%
        {Khanna07}
\bibfield{author}{\bibinfo{person}{Sanjeev Khanna}, \bibinfo{person}{Keshav
  Kunal}, {and} \bibinfo{person}{Benjamin~C. Pierce}.}
  \bibinfo{year}{2007}\natexlab{}.
\newblock \showarticletitle{A Formal Investigation of Diff3}. In
  \bibinfo{booktitle}{\emph{FSTTCS 2007: Foundations of Software Technology and
  Theoretical Computer Science}}, \bibfield{editor}{\bibinfo{person}{V.~Arvind}
  {and} \bibinfo{person}{Sanjiva Prasad}} (Eds.). \bibinfo{publisher}{Springer
  Berlin Heidelberg}, \bibinfo{address}{Berlin, Heidelberg},
  \bibinfo{pages}{485--496}.
\newblock
\showISBNx{978-3-540-77050-3}


\bibitem[\protect\citeauthoryear{Kleppmann, Mulligan, Gomes, and
  Beresford}{Kleppmann et~al\mbox{.}}{2021}]%
        {Kleppmann21}
\bibfield{author}{\bibinfo{person}{Martin Kleppmann},
  \bibinfo{person}{Dominic~P. Mulligan}, \bibinfo{person}{Victor B.~F. Gomes},
  {and} \bibinfo{person}{Alastair Beresford}.} \bibinfo{year}{2021}\natexlab{}.
\newblock \showarticletitle{A highly-available move operation for replicated
  trees}.
\newblock \bibinfo{journal}{\emph{IEEE Transactions on Parallel and Distributed
  Systems}} (\bibinfo{year}{2021}), \bibinfo{pages}{1--1}.
\newblock
\urldef\tempurl%
\url{https://doi.org/10.1109/TPDS.2021.3118603}
\showDOI{\tempurl}


\bibitem[\protect\citeauthoryear{K\"{o}lling, Brown, and Altadmri}{K\"{o}lling
  et~al\mbox{.}}{2015}]%
        {Kolling15}
\bibfield{author}{\bibinfo{person}{Michael K\"{o}lling}, \bibinfo{person}{Neil
  C.~C. Brown}, {and} \bibinfo{person}{Amjad Altadmri}.}
  \bibinfo{year}{2015}\natexlab{}.
\newblock \showarticletitle{Frame-Based Editing: Easing the Transition from
  Blocks to Text-Based Programming}. In \bibinfo{booktitle}{\emph{Proceedings
  of the Workshop in Primary and Secondary Computing Education}} (London,
  United Kingdom) \emph{(\bibinfo{series}{WiPSCE '15})}.
  \bibinfo{publisher}{Association for Computing Machinery},
  \bibinfo{address}{New York, NY, USA}, \bibinfo{pages}{29–38}.
\newblock
\showISBNx{9781450337533}
\urldef\tempurl%
\url{https://doi.org/10.1145/2818314.2818331}
\showDOI{\tempurl}


\bibitem[\protect\citeauthoryear{Litt, van Hardenberg, and Orion}{Litt
  et~al\mbox{.}}{2020}]%
        {Cambria}
\bibfield{author}{\bibinfo{person}{Geoffrey Litt}, \bibinfo{person}{Peter van
  Hardenberg}, {and} \bibinfo{person}{Henry Orion}.}
  \bibinfo{year}{2020}\natexlab{}.
\newblock \bibinfo{title}{Project Cambria: Translate your data with lenses}.
\newblock
\newblock
\urldef\tempurl%
\url{https://www.inkandswitch.com/cambria.html}
\showURL{%
\tempurl}


\bibitem[\protect\citeauthoryear{Malik}{Malik}{1997}]%
        {Envy}
\bibfield{author}{\bibinfo{person}{Vikas Malik}.}
  \bibinfo{year}{1997}\natexlab{}.
\newblock \bibinfo{title}{Smalltalk Envy}.
\newblock
\newblock
\urldef\tempurl%
\url{http://www.faqs.org/faqs/smalltalk/ENVY-faq/}
\showURL{%
\tempurl}


\bibitem[\protect\citeauthoryear{Mimram and Di~Giusto}{Mimram and
  Di~Giusto}{2013}]%
        {PatchTheory}
\bibfield{author}{\bibinfo{person}{Samuel Mimram} {and} \bibinfo{person}{Cinzia
  Di~Giusto}.} \bibinfo{year}{2013}\natexlab{}.
\newblock \showarticletitle{A Categorical Theory of Patches}.
\newblock \bibinfo{journal}{\emph{Electronic Notes in Theoretical Computer
  Science}}  \bibinfo{volume}{298} (\bibinfo{date}{Nov} \bibinfo{year}{2013}),
  \bibinfo{pages}{283–307}.
\newblock
\showISSN{1571-0661}
\urldef\tempurl%
\url{https://doi.org/10.1016/j.entcs.2013.09.018}
\showDOI{\tempurl}


\bibitem[\protect\citeauthoryear{Omar, Voysey, Chugh, and Hammer}{Omar
  et~al\mbox{.}}{2019}]%
        {HazelnutLive19}
\bibfield{author}{\bibinfo{person}{Cyrus Omar}, \bibinfo{person}{Ian Voysey},
  \bibinfo{person}{Ravi Chugh}, {and} \bibinfo{person}{Matthew~A. Hammer}.}
  \bibinfo{year}{2019}\natexlab{}.
\newblock \showarticletitle{Live Functional Programming with Typed Holes}.
\newblock \bibinfo{journal}{\emph{{PACMPL}}} \bibinfo{volume}{3},
  \bibinfo{number}{{POPL}} (\bibinfo{year}{2019}).
\newblock
\urldef\tempurl%
\url{https://doi.org/10.1145/3290327}
\showDOI{\tempurl}


\bibitem[\protect\citeauthoryear{Omar, Voysey, Hilton, Aldrich, and
  Hammer}{Omar et~al\mbox{.}}{2017}]%
        {Hazelnut17}
\bibfield{author}{\bibinfo{person}{Cyrus Omar}, \bibinfo{person}{Ian Voysey},
  \bibinfo{person}{Michael Hilton}, \bibinfo{person}{Jonathan Aldrich}, {and}
  \bibinfo{person}{Matthew~A. Hammer}.} \bibinfo{year}{2017}\natexlab{}.
\newblock \showarticletitle{{Hazelnut: A Bidirectionally Typed Structure Editor
  Calculus}}. In \bibinfo{booktitle}{\emph{44th {ACM} {SIGPLAN} Symposium on
  Principles of Programming Languages ({POPL} 2017)}}.
\newblock


\bibitem[\protect\citeauthoryear{Oster, Molli, Urso, and Imine}{Oster
  et~al\mbox{.}}{2006}]%
        {Oster06}
\bibfield{author}{\bibinfo{person}{Gerald Oster}, \bibinfo{person}{Pascal
  Molli}, \bibinfo{person}{Pascal Urso}, {and} \bibinfo{person}{Abdessamad
  Imine}.} \bibinfo{year}{2006}\natexlab{}.
\newblock \showarticletitle{Tombstone Transformation Functions for Ensuring
  Consistency in Collaborative Editing Systems}. In
  \bibinfo{booktitle}{\emph{2006 International Conference on Collaborative
  Computing: Networking, Applications and Worksharing}}.
  \bibinfo{pages}{1--10}.
\newblock
\urldef\tempurl%
\url{https://doi.org/10.1109/COLCOM.2006.361867}
\showDOI{\tempurl}


\bibitem[\protect\citeauthoryear{Resnick, Maloney, Monroy-Hern\'{a}ndez, Rusk,
  Eastmond, Brennan, Millner, Rosenbaum, Silver, Silverman, and Kafai}{Resnick
  et~al\mbox{.}}{2009}]%
        {Resnick09}
\bibfield{author}{\bibinfo{person}{Mitchel Resnick}, \bibinfo{person}{John
  Maloney}, \bibinfo{person}{Andr\'{e}s Monroy-Hern\'{a}ndez},
  \bibinfo{person}{Natalie Rusk}, \bibinfo{person}{Evelyn Eastmond},
  \bibinfo{person}{Karen Brennan}, \bibinfo{person}{Amon Millner},
  \bibinfo{person}{Eric Rosenbaum}, \bibinfo{person}{Jay Silver},
  \bibinfo{person}{Brian Silverman}, {and} \bibinfo{person}{Yasmin Kafai}.}
  \bibinfo{year}{2009}\natexlab{}.
\newblock \showarticletitle{Scratch: Programming for All}.
\newblock \bibinfo{journal}{\emph{Commun. ACM}} \bibinfo{volume}{52},
  \bibinfo{number}{11} (\bibinfo{date}{Nov.} \bibinfo{year}{2009}),
  \bibinfo{pages}{60–67}.
\newblock
\showISSN{0001-0782}
\urldef\tempurl%
\url{https://doi.org/10.1145/1592761.1592779}
\showDOI{\tempurl}


\bibitem[\protect\citeauthoryear{Ressel, Nitsche-Ruhland, and
  Gunzenh\"{a}user}{Ressel et~al\mbox{.}}{1996}]%
        {Ressel96}
\bibfield{author}{\bibinfo{person}{Matthias Ressel}, \bibinfo{person}{Doris
  Nitsche-Ruhland}, {and} \bibinfo{person}{Rul Gunzenh\"{a}user}.}
  \bibinfo{year}{1996}\natexlab{}.
\newblock \showarticletitle{An Integrating, Transformation-Oriented Approach to
  Concurrency Control and Undo in Group Editors}. In
  \bibinfo{booktitle}{\emph{Proceedings of the 1996 ACM Conference on Computer
  Supported Cooperative Work}} (Boston, Massachusetts, USA)
  \emph{(\bibinfo{series}{CSCW '96})}. \bibinfo{publisher}{Association for
  Computing Machinery}, \bibinfo{address}{New York, NY, USA},
  \bibinfo{pages}{288–297}.
\newblock
\showISBNx{0897917650}
\urldef\tempurl%
\url{https://doi.org/10.1145/240080.240305}
\showDOI{\tempurl}


\bibitem[\protect\citeauthoryear{Sandewall}{Sandewall}{1978}]%
        {Sandewall78}
\bibfield{author}{\bibinfo{person}{Erik Sandewall}.}
  \bibinfo{year}{1978}\natexlab{}.
\newblock \showarticletitle{Programming in an Interactive Environment: The
  ``Lisp'' Experience}.
\newblock \bibinfo{journal}{\emph{ACM Comput. Surv.}} \bibinfo{volume}{10},
  \bibinfo{number}{1} (\bibinfo{date}{March} \bibinfo{year}{1978}),
  \bibinfo{pages}{35–71}.
\newblock
\showISSN{0360-0300}
\urldef\tempurl%
\url{https://doi.org/10.1145/356715.356719}
\showDOI{\tempurl}


\bibitem[\protect\citeauthoryear{s.r.o.}{s.r.o.}{2021}]%
        {MPS}
\bibfield{author}{\bibinfo{person}{JetBrains s.r.o.}}
  \bibinfo{year}{2021}\natexlab{}.
\newblock \bibinfo{title}{MPS: The Domain-Specific Language Creator}.
\newblock
\newblock
\urldef\tempurl%
\url{https://www.jetbrains.com/mps/}
\showURL{%
\tempurl}


\bibitem[\protect\citeauthoryear{Systems}{Systems}{2015}]%
        {Gemstone}
\bibfield{author}{\bibinfo{person}{GemTalk Systems}.}
  \bibinfo{year}{2015}\natexlab{}.
\newblock \bibinfo{title}{Class versions and Instance Migration}.
\newblock
\newblock
\urldef\tempurl%
\url{https://downloads.gemtalksystems.com/docs/GemStone64/3.2.x/GS64-ProgGuide-3.2/10-ClassHistory.htm}
\showURL{%
\tempurl}


\bibitem[\protect\citeauthoryear{Teitelbaum and Reps}{Teitelbaum and
  Reps}{1981}]%
        {Teitelbaum81}
\bibfield{author}{\bibinfo{person}{Tim Teitelbaum} {and}
  \bibinfo{person}{Thomas Reps}.} \bibinfo{year}{1981}\natexlab{}.
\newblock \showarticletitle{The Cornell Program Synthesizer: A Syntax-Directed
  Programming Environment}.
\newblock \bibinfo{journal}{\emph{Commun. ACM}} \bibinfo{volume}{24},
  \bibinfo{number}{9} (\bibinfo{date}{Sept.} \bibinfo{year}{1981}),
  \bibinfo{pages}{563–573}.
\newblock
\showISSN{0001-0782}
\urldef\tempurl%
\url{https://doi.org/10.1145/358746.358755}
\showDOI{\tempurl}


\bibitem[\protect\citeauthoryear{Trenouth}{Trenouth}{1991}]%
        {Trenouth91}
\bibfield{author}{\bibinfo{person}{J. Trenouth}.}
  \bibinfo{year}{1991}\natexlab{}.
\newblock \showarticletitle{{A Survey of Exploratory Software Development}}.
\newblock \bibinfo{journal}{\emph{Comput. J.}} \bibinfo{volume}{34},
  \bibinfo{number}{2} (\bibinfo{date}{01} \bibinfo{year}{1991}),
  \bibinfo{pages}{153--163}.
\newblock
\showISSN{0010-4620}
\urldef\tempurl%
\url{https://doi.org/10.1093/comjnl/34.2.153}
\showDOI{\tempurl}
\showeprint{https://academic.oup.com/comjnl/article-pdf/34/2/153/1400604/340153.pdf}


\end{thebibliography}

\appendix

\section{Projection and retraction rules}
\label{rules}

Equal edits cancel out:

$
\begin{tikzcd}[column sep = large]
\hole \ar[d, swap, "x"] \ar[r, "x"]
& \hole \ar[d, dashed, "\Id"]\\
\hole \ar[r, dashed, swap, "\Id"]
& \hole
\end{tikzcd}
\quad
\begin{tikzcd}[column sep = large]
\hole \ar[d, dashed, swap, "\Id"] \ar[r, "x"]
& \hole \ar[d, "x"]\\
\hole \ar[r, dashed, swap, "x"]
& \hole
\end{tikzcd}
$

$\Id$ is a fixpoint:

$
\begin{tikzcd}
\hole  \ar[d, swap, "\Id"] \ar[r, "x"]
&\hole  \ar[d, "\Id"]\\
\hole  \ar[r, swap, "x"]
&\hole
\end{tikzcd}
\quad
\begin{tikzcd}
\hole  \ar[d, swap, "x"] \ar[r, "\Id"]
&\hole  \ar[d, "x"]\\
\hole  \ar[r, swap, "\Id"]
&\hole
\end{tikzcd}
$

$\Conv$ is overridden by conflicting changes:

$
\begin{tikzcd}[column sep = large]
\hole \ar[d, swap, "x"] \ar[r, "{\Conv[i, u]}"]
& \hole \ar[d, "x"] \\
\hole \ar[r, swap, "\Id"]
& \hole\\
\end{tikzcd}
$
\quad
\begin{scriptsize}
$
\begin{cases}
x = \Conv[i, t] \land t \neq u\\
x = \Move[i, \_]
\end{cases}
$
\end{scriptsize}

$\Conv$ passes through independent edits:

$
\begin{tikzcd}[column sep = large]
\hole \ar[d, swap, "x"] \ar[r, "{\Conv[i, t]}"]
& \hole \ar[d, "x"] \\
\hole \ar[r, swap, "{\Conv[i, t]}"]
& \hole\\
\end{tikzcd}
$
\quad
\begin{scriptsize}
$
i \neq j
\ \land \
\begin{cases}
x = \Conv[j, \_]\\
x = \Move[j, k] \land i \neq k\\
\end{cases}
$
\end{scriptsize}

$\Ins$ doesn't affect leftward edits:

$
\begin{tikzcd}[column sep=large]
\hole  \ar[d, swap, "x"] \ar[r, "{\Ins[i, t, p]}"]
&\hole  \ar[d, "x"]\\
\hole  \ar[r,  swap, "{\Ins[i, t, p]}"]
&\hole
\end{tikzcd}
\quad
\begin{tikzcd}[column sep=large]
\hole  \ar[d, swap, "{\Ins[i, t, p]}"] \ar[r, "x"]
&\hole  \ar[d, "{\Ins[i, t, p]}"]\\
\hole  \ar[r,  swap, "x"]
&\hole
\end{tikzcd}
$
\quad
\begin{scriptsize}
$
i > j
\ \land \
\begin{cases}
x = \Conv[j, \_]\\
x = \Move[j, k] \land i > k\\
\end{cases}
$
\end{scriptsize}

$\Ins$ shifts non-leftward edits rightward:

$
\begin{tikzcd}[column sep=large]
\hole  \ar[d, swap, "x"] \ar[r, "{\Ins[i, t, p]}"]
&\hole  \ar[d, "y"]\\
\hole  \ar[r,  swap, "{\Ins[i, t, p]}"]
&\hole
\end{tikzcd}
\quad
\begin{tikzcd}[column sep=large]
\hole  \ar[d, swap, "{\Ins[i, t, p]}"] \ar[r, "x"]
&\hole  \ar[d, "{\Ins[i, t, p]}"]\\
\hole  \ar[r,  swap, "y"]
&\hole
\end{tikzcd}
$
\quad
\begin{scriptsize}
$
i \leq j
\ \land \
\begin{cases}
x = \Conv[j, u] \land y = \Conv[j + 1, u]\\
x = \Move[j, k] \land y = \Move[j+1, k] \land i > k\\
x = \Move[k, j] \land y = \Move[k, j+1] \land i > k\\
x = \Move[j, k] \land y = \Move[j+1, k+1] \land i \leq k\\
\end{cases}
$
\end{scriptsize}

$
\begin{tikzcd}[column sep=large]
\hole  \ar[d, swap, "{\Ins[j, u, q]}"] \ar[r, "{\Ins[i, t, p]}"]
&\hole  \ar[d, "{\Ins[j+1, u, q]}"]\\
\hole  \ar[r,  swap, "{\Ins[i, t, p]}"]
&\hole
\end{tikzcd}
$
\quad
\begin{scriptsize}
$
i \leq j
\ \land \
p \neq q
$
\end{scriptsize}

$
\begin{tikzcd}[column sep=large]
\hole  \ar[d, swap, "{\Ins[j, u, q]}"] \ar[r, "{\Ins[i, t, p]}"]
&\hole  \ar[d, "{\Ins[j, u, q]}"]\\
\hole  \ar[r,  swap, "{\Ins[i + 1, t, p]}"]
&\hole
\end{tikzcd}
$
\quad
\begin{scriptsize}
$
i > j
\ \land \
p \neq q
$
\end{scriptsize}

Cannot retract an edit at a location through the insert of that location. Not even an insert at that location, because that would not be reversible by a projection.

$
\begin{tikzcd}[column sep=large]
\hole  \ar[d, dashed, "/" marking] \ar[r, "{\Ins[i, \_, p]}"]
&\hole  \ar[d, "x"]\\
\hole  \ar[r, dashed, "/" marking]
&\hole
\end{tikzcd}
$
\quad
\begin{scriptsize}
$
\begin{cases}
x = \Ins[i, \_, q] \land p \neq q\\
x = \Conv[i, \_]\\
x = \Move[i, \_]\\
x = \Move[\_, i]\\
\end{cases}
$
\end{scriptsize}

$\Move$ moves source to target:

$
\begin{tikzcd}[column sep=large]
\hole  \ar[d, swap, "x"] \ar[r, "{\Move[i, j]}"]
&\hole  \ar[d, "y"]\\
\hole  \ar[r,  swap, "{\Move[i, j]}"]
&\hole
\end{tikzcd}
\quad
\begin{tikzcd}[column sep=large]
\hole  \ar[d, swap, "{\Move[i, j]}"] \ar[r, "x"]
&\hole  \ar[d, "{\Move[i, j]}"]\\
\hole  \ar[r,  swap, "y"]
&\hole
\end{tikzcd}
$
\quad
\begin{scriptsize}
$
\begin{cases}
x = \Conv[j, t] \land y = \Conv[i, t]\\
x = \Move[j, k] \land y = \Move[i, k] \land i \neq k\\
x = \Move[k, j] \land y = \Move[k, i] \land i \neq k\\
\end{cases}
$
\end{scriptsize}

Conflicting $\Move$ edits delete the overridden Move's source:

$
\begin{tikzcd}[column sep=large]
\hole  \ar[d, swap, "{\Move[i, k]}"] \ar[r, "{\Move[i, j]}"]
&\hole  \ar[d, "{\Move[i, k]}"]\\
\hole  \ar[r,  swap, "{\Conv[j, \del]}"]
&\hole
\end{tikzcd}
$
\quad
\begin{scriptsize}
$j \neq k$
\end{scriptsize}

Can't project $\Conv$ through $\Move$ to its target, because it would retract to the source of the Move.

$
\begin{tikzcd}[column sep=large]
\hole  \ar[d, swap, "{\Conv[i, \_]}"] \ar[r, "{\Move[i, j]}"]
&\hole  \ar[d, dashed, "\Id"]\\
\hole  \ar[r, dashed, swap, "{\Move[i, j]}"]
&\hole
\end{tikzcd}
$

$\Move$ passes through edits not to target or source:

$
\begin{tikzcd}[column sep=large]
\hole  \ar[d, swap, "x"] \ar[r, "{\Move[i, j]}"]
&\hole  \ar[d, "x"]\\
\hole  \ar[r,  swap, "{\Move[i, j]}"]
&\hole
\end{tikzcd}
$
\quad
\begin{scriptsize}
$
i \neq k
\ \land \
j \neq k
\ \land \
\begin{cases}
x = \Conv[k, \_]\\
x = \Move[k, l] \land i \neq l \land j \neq l\\
\end{cases}
$
\end{scriptsize}


\end{document}